\documentclass{aa}  

\usepackage[position=top]{subfig}
\usepackage{graphicx}
\usepackage{csquotes}
\usepackage{txfonts}
\usepackage{svg}
\usepackage[T1]{fontenc}
\usepackage[normalem]{ulem}
\usepackage{xcolor}
\usepackage{gensymb}
\usepackage[colorlinks=true,linkcolor=blue,allcolors=blue]{hyperref}
\usepackage[capitalise]{cleveref}
\definecolor{blackberry}{HTML}{8D1D75}
\definecolor{lightblue}{rgb}{0.1,0.5,0.89}

\usepackage{float}
\newcommand\bluesout{\bgroup\markoverwith{\textcolor{teal}{\rule[0.5ex]{2pt}{0.4pt}}}\ULon}

\begin{document}

\title{Non-spherical Cows: Introducing the Asphericity Parameter as a Measure of Accretion Geometry}
\titlerunning{Inflow asphericity}

   \author{Benjamin A. Seidel\inst{\ref{inst:usm}}\thanks{E-mail: bseidel@usm.uni-muenchen.de}
           \and
           Rhea-Silvia Remus\inst{\ref{inst:usm},\ref{inst:swin}}
           \and
           Lucas C. Kimmig\inst{\ref{inst:usm}}
           \and
           Lucas M. Valenzuela\inst{\ref{inst:usm}}
           \and
           Klaus Dolag\inst{\ref{inst:usm},\ref{inst:mpa}}
          }

   \institute{Universitäts-Sternwarte, Fakultät für Physik, Ludwig-Maximilians Universität München, Scheinerstr.1, 81679 München, Germany\label{inst:usm}
   \and 
   Centre for Astrophysics and Supercomputing, Swinburne University of Technology, Hawthorn VIC 3122, Australia\label{inst:swin}
   \and
   Max-Planck-Institute for Astrophysics, Karl-Schwarzschild-Str.\ 1, 85748 Garching, Germany\label{inst:mpa}
             }

   \date{}

 
  \abstract
   {The outer regions of galactic halos represent the bridge connecting internal processes within the galaxy to the larger surrounding cosmic web. The gas in this bridge region is shaped by the competing processes of cold inflows from the web and hot ejecta from feedback of supernovae or an active galactic nucleus. Similarly, the gas around galaxy clusters characterizes the balance between inflows and outflows. To study this connection, we introduce a new parameter for quantifying the geometrical configuration of the flow field connecting structures to the cosmic web, the asphericity parameter. This inflow asphericity is based on a spherical harmonics decomposition of the inflow at the virial boundary of the halo. It can be computed using both the linear and the logarithmic inflow field. To validate this parameter we apply it to both an extensive toy model set and to simulated haloes from the Magneticum simulations. We find the linear asphericity to be a tracer of the total power of the non-spherical inflow and the total anisotropy. On the other hand, the logarithmic asphericity traces the covering fraction of inflows at the surface and is highly sensitive to regions with zero inflow (regions that are dominated by outflow). Thus the asphericity of the flow field is a powerful tool to simultaneously study the geometry of in- and outflows in numerical simulations.}

   \maketitle
%
\section{Introduction}
Galaxies, groups, and clusters of galaxies evolve from the collapse of primordial perturbations in the cosmic density field. As these perturbations are never perfectly spherical, their gravitational collapse also happens non-spherically. The resulting tidal fields give rise to a number of dynamical and morphological phenomena such as galactic spins \citep{porciani2002,neyrinck2020} and on larger scales to the cosmic web with its elongated filaments and sheets \citep[e.g.][]{bond1996,hidding2014}.

Quantifying the morphology and topology of the cosmic web and its filaments has been a major goal of both theoretical \citep{sousbie2011,kuchner2021} and observational \citep{hoang2023,euclidcollaboration2025} efforts since its first theoretical conception by \citet{zeldovich1970}. These filaments connect the massive clusters and superclusters that sit at the nodes of the web and have been found to affect several internal properties of these structures \citep{gouin2021,haggar2024}. 

From the perspective of galaxy formation especially the influence of the cosmic web on smaller scales is a topic of major interest. First studies connecting galaxy properties to various parameters quantifying the position in the cosmic web show promising correlations between internal and external physics: Multiple studies, both observational \citep{alpaslan2016,malavasi2017,kraljic2018,hoosain2024} and numerical \citep{bulichi2024,hasan2024}, found galaxies at the cores of filaments to be biased towards higher stellar masses while the expected star formation rate increases with the distance to the spine. This is also in agreement with theoretical expectations from excursion set theory \citep{musso2018}, as in the vicinity of saddle points, accretion is expected to be influenced by the large-scale flows distorting the local velocity field. 

A key issue with connecting the cosmological environment to galaxy evolution is that an unambiguous definition of cosmic structures is difficult to devise \citep{libeskind2018}, due to the complex, multi-scalar nature of the web. In simulations, where one has access to the full density fields, this can be mitigated to a degree with multi-scalar structure identification algorithms like NEXUS and NEXUS+ \citep{cautun2013}. This is, however, not feasible for observations, where the density environment of a galaxy can typically only be constrained only on larger scales by its surrounding galaxy distribution \citep[e.g.][]{alpaslan2016,kraljic2018,euclidcollaboration2025} and on small scales by trying to map the circumgalactic medium and cold gas components in the outskirts of a given galaxy. The intermediate scale gas distribution is currently only sparsely accessible to observations in field haloes, for example.

Additionally from modelling galaxy evolution in N-body simulations it became clear that the story of galactic gas accretion is more complex than the growth of dark matter haloes from purely cosmological accretion. Simulation models like the semi-analytic Millennium model \citep{croton2006} diverged at the low mass end of the stellar mass function at high-z and at the high-mass end at low-z \citep{bell2003,li2009}. This implicated that galaxy scale outflows play an important role as a regulation mechanism for accretion and thus also star formation. Direct evidence for these regulating gas flows has since been found observationally \citep{roberts-borsani2019} and therefore most modern hydrodynamical cosmological simulations employ a combination of AGN and star-formation driven outflows from sub-grid modelling that are able to bring the galaxy stellar mass functions closer to the observations \citep{schaye2015,dave2019,nelson2021,schaye2023,schaye2025,dolag2025}. Nevertheless, these models come in many different flavours and the large-scale gas environment of galaxies has been shown to be sensitive to the specific implementation of the model or even parameter choices within a given model \citep{ayromlou2023}. From a theoretical perspective, it is therefore a valuable exercise to consider the gas flow fields that shape these environments directly and study how the flows are connected to the internal galaxy evolution.

As the cosmic web evolves with time \citep{cautun2014} and galaxies evolve in terms of their stellar and gas content, the balance 
between inflow from the environment and outflows can be expected to change as well. A baseline equilibrium state is given by the so-called 'bathtub models', which directly relate the net gas inflow rate to the star-formation rate \citep{bouche2010,dave2012,lilly2013}. Recent numerical analysis showed that, while galaxies generally follow these equilibrium relations at early times ($z>2$), there is a transition towards late times with fewer galaxies following a simple relation between flow rate and star formation \citep{fortune2025}. This is also consistent with the finding that massive quenched galaxies accrete relatively isotropically at high-redshift until they quench and disrupt their environment significantly \citep{kimmig2025a}. This disruption, in combination with a general hot halo build-up for massive galaxies can effectively cut them off from the cold gas inflows from the cosmic web \citep{dekel2009}, which arise as anisotropies in the flow field around the galaxy. From numerical simulations \citet{remus2023} showed that such hot haloes can already exist as early as $z=4.2$ and a first observation of a hot X-ray halo was recently published \citep{bogdan2026}. 

The morphology of the flow field around a galaxy is therefore dependent on many complex processes from cosmic web evolution to stellar and AGN outflows. In this work we aim to quantify how (an)-isotropic the inflow field is for a given halo and to this end present and test a new parameter, the \textbf{inflow asphericity} ($\zeta$) of a given simulated halo. This parameter is computed on an input inflow field measured at a certain boundary surface (e.g. a sphere with one virial radius) centred on the halo of interest. 

In previous work we already demonstrated the flow fields on the outskirts of haloes to be vastly diverse from a morphological perspective, nevertheless exhibiting a clear trend of inflows being more anisotropically distributed than the more spherically symmetric outflows \citep{seidel2025}. In that paper we employed a spherical harmonics decomposition of the flow field to measure how isotropic the flow field structure is, which gave rise to the inflow asphericity parameter. First exploratory studies employing this parameter started connecting the geometry of the inflow field to both the shapes \citep{valenzuela2024} and the star formation histories \citep{fortune2025} of galaxies, making it a promising quantity to understand galaxy-environment connections better from a theoretical perspective. In this work we introduce this inflow asphericity parameter and explore and verify its behaviour in different idealized and simulated setups.

This paper is structured as follows: 
In \cref{Sec:Methods}, we introduce the basic methods we used to compute the asphericity parameter from a given flow field, as well as the simulations we employed to test this parameter. In \cref{sec:testcases} we introduce idealized test cases to assess how the asphericity parameter corresponds to basic properties of the inflow field. In \cref{sec:Simtest} we demonstrate the spherical harmonics decomposition and the computation of the asphericity parameter using simulated maps from hydrodynamical simulations.

\section{Introducing the asphericity parameter} \label{Sec:Methods}
  \begin{figure*}[!htbp]
\centering
\includegraphics[width=\textwidth]{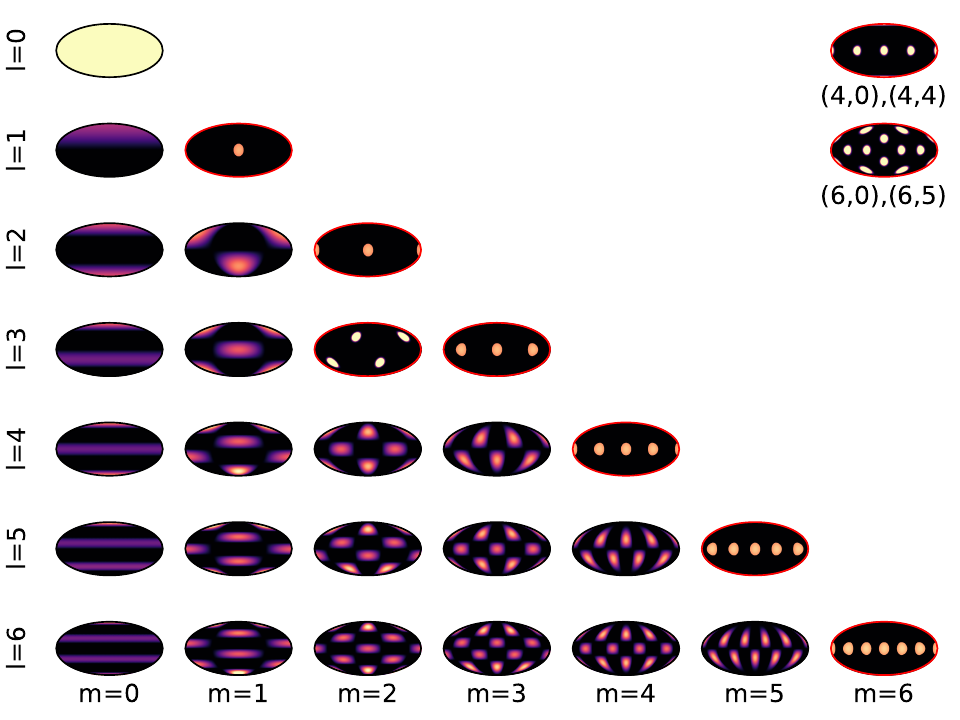}
\caption{Spherical harmonics as surface maps. The row corresponds to the $l$-degree of the spherical harmonic, while the column indicates the order $m$ (from 0 to 6). We show the planar test maps (see \cref{sec:testcases}) at the positions of the sectoral harmonics with the corresponding $l,m$ highlighted in red.  We additionally show the platonic solid test maps (see \cref{platonics}): The Tetrahedron configuration is shown at the position of the pure harmonic $Y_{32}$ which has tetrahedral geometry. The two additional platonic solids, octahedron and icosahedron, which can be represented by linear combinations of spherical harmonics in the top right columns, with their corresponding $(l,m)$ values given below.}
\label{triangle}
\end{figure*}
The idea of the asphericity parameter is to quantify how spherically symmetrical the flow field around a given halo is, or, conversely, how much large-scale variation there is in the inflow field. In the idealized case of perfectly smooth and purely spherical accretion, the inflow will be the same in every direction from the perspective of an observer located at the centre of the halo. Deviations from this can have multiple sources: Filaments and infalling secondary haloes will create accretion hotspots, angular directions along which much more mass gets accreted compared to the average accretion rate. Throughout this work we will refer to this mode of accretion as the "filamentary component". Since major satellite mergers can be expected to generally align with the filamentary structure \citep[e.g.][]{arora2025}, this nomenclature is reasonable. Outflows on the other hand can, if they propagate out far enough, stop accretion entirely in the directions they are launched. Our goal is therefore to quantify these deviations from isotropic accretion as a measure of the evolution of the large scale structure a halo is connected to, as well as the internal processes of the halo such as star formation and AGN activity.
\subsection{Spherical Harmonics Decomposition}
 In order to quantify the geometry of the inflow field, we perform a spherical harmonics decomposition. Spherical harmonics decompositions are routinely employed in cosmology to study angular power spectra from the cosmic microwave background \citep[e.g.][]{akrami2020c} or galaxy distributions \citep{tessore2025}. Analogous to a Fourier decomposition the general idea is to break a given function down into a sum of oscillating basis function with a well defined degree $l$, analogous to the frequency in the Fourier method, and an order $m$. 
 
 We decompose the inflow function $f(\theta,\phi)$ using a set of orthogonal spherical harmonics functions
 \begin{equation}
    f(\theta,\phi)=\sum_{l=0}^{l_\mathrm{max}}\sum_{m=-l}^{m=l}a_{l,m}Y_{l,m}(\theta,\phi) .
 \end{equation} From the coefficients $a_{l,m}$ one can compute the power spectrum $[C_l]$ via 
 \begin{equation}
 C_l=\frac{1}{2l+1}\sum_{m=-l}^l |a_{l,m}|^2 .
 \end{equation} The total power in each $l$-mode is given by $(2l+1)C_l$. The decompositions in this work are computed using the \textit{healpy} wrapper \citep{Zonca2019} for healpix tessellated maps \citep{gorski2005a}.
 
 \cref{triangle} shows these spherical harmonics as Mollweide projected surface maps. We deliberately normalize these maps in such a way that they are comparable to the test maps discussed in \cref{sec:testcases}, namely instead of highlighting zones of positive and negative values, we focus on the positive regions here as they will be the proxies for our zones of inflow. Where there is a 1:1 correspondence between harmonics and synthetic test cases, we show the test maps at the appropriate position with a red frame. Additionally two of the polyhedral setups using the platonic solids (see \cref{platonics}) can be obtained by appropriately combining two spherical harmonics. We therefore show the Octahedron and Icosahedron setup in the upper right of the figure, giving the harmonics they can be obtained from. 
 \subsection{Asphericity}
 Techniques employing either projected or spherical harmonics decompositions have been used in the past to identify preferred directions of infall for satellite galaxies \citep{libeskind2011,arora2025}, to measure the degree of filamentary to isotropic accretion in isolated cluster simulations \citep{valles-perez2020}, and to study the correlation of the 2D-azimuthal symmetries in gas and DM to various halo properties \citep{gouin2022}. In this work we calculate the aspherical excess on the sphere. This excess is given by the sum of all multipole moments higher than the monopole normalized to the monopole power, truncating the sum at the 9-th degree to ideally capture the large-scale anisotropic inflow modes:
 \begin{equation}
    \zeta=\frac{1}{C_0}\sum_{l=1}^{9} (2l+1)C_l.
    \label{equation:asphericity}
\end{equation}
In principle the choice of cut-off is arbitrary, we arrive at $l=9$ for the following reasons: For $l=9$ a typical angular scale of fluctuations is $\pi/9$ or $20\degree$. At the virial radius of a massive galaxy cluster (assuming $r_\mathrm{vir}=1$ Mpc) this corresponds to a physical scale of $\approx 0.35 $ Mpc, which can be expected to capture the physical width of simulated cosmological filaments at all redshifts \citep[e.g.][]{yang2025}. Additionally, using the combined power spectra of 2000 haloes from the Magneticum simulations (see \cref{sec:methods:pecflow}) we found that this harmonic degree marks a transition point between the large-scale power spectrum and the smaller scale features of a typical simulated inflow map (see \cref{appendix:aggregate}).

A more homogeneous field has less power in the non-isotropic angular modes, thus lowering $\zeta$ with a perfectly homogenous inflow field having a $\zeta$ of 0. For a more complex inflow field the exact value of $\zeta$ will depend on the spatial configuration (for example filament thickness and alignment), so to disentangle the exact structure the full spectrum of the spherical harmonics is necessary (see the discussion in \cref{appendix:connectivity}).
\subsection{Logarithmic versus linear asphericity}
We introduce two complementary methods to compute the large-angle fluctuations in the flow field: From \cref{calib} it can be seen that the computed inflow values around simulated haloes span a wide dynamical range. Therefore the purely linear maps will be dominated by the regions where the inflow is maximal. A way to reduce the dynamical range of a given inflow map is to instead analyse the spatial distribution in log space instead.  As will be demonstrated in the following, computing the asphericity on the linear flow field, denoted $\zeta_\mathrm{lin}$ hereafter, is most sensitive to the dominant regions of inflow, where the flow is several orders of magnitude above the inflow averaged over the spherical surface. Consequently it serves to measure the amount of inflow through filaments and infalling galaxies, where densities are the highest. Computing the asphericity on the logged map on the other hand ($\zeta_\mathrm{log}$) is more sensitive to the overall covering fraction of the inflowing component on the virial surface. This makes this variant of the asphericity more sensitive to outflows driven by the feedback processes inside the halo, as will be demonstrated in the next section.

\section{Idealized test cases}
\label{sec:testcases}
\subsection{Test setup}
The inflow maps of simulated haloes are inherently complex and a superposition of many components both in real and in spherical harmonics space. To better illustrate how the asphericity parameter works, we set up a set of idealized test cases.
The basic idea of these test setups is to break the complex inflow maps down to three components:
\begin{itemize}
    \item A filamentary component, represented by $N_\mathrm{fil}$ hotspots on the sphere of solid angle extent $d$ and a total inflow power $j_\mathrm{fil}$
    \item An isotropic component with a base inflow level $j_0$
    \item Holes in the isotropic component (i.e. regions with zero value) representing regions of outflow, characterized by either a cone or a bi-cone with a total solid angle $\Omega_\mathrm{outflow}$. We parametrize this by the surface ratio $f_\mathrm{outflow}$, the fraction of the spherical surface covered by the outflow.
\end{itemize}
\begin{figure}
\centering
\includegraphics[width=\columnwidth]{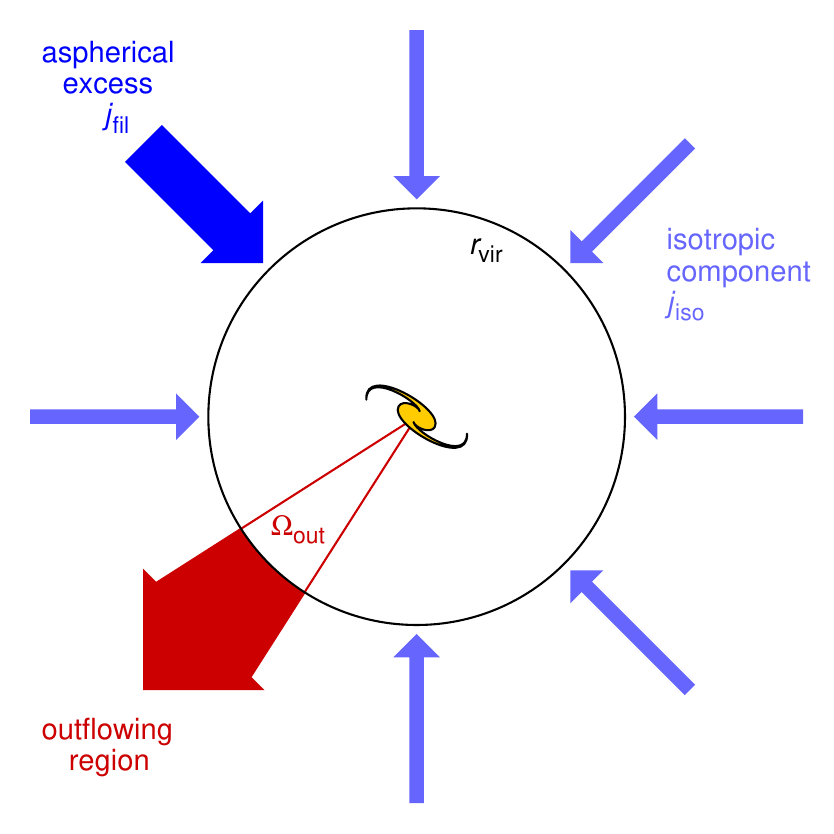}
\caption{Schematic of the toy model setup.}
\label{toy}
\end{figure}

\cref{toy} visualizes these three components.
We adapted a Gaussian profile for the intensity distribution within each filament (the cross-section corresponds to the full width at half maximum (FWHM) here).
As the default FWHM we chose $0.2$ rad, which corresponds to a physical filament width of $\approx0.2r_\mathrm{vir}$ assuming the flow field is measured at the virial radius. The goal of this idealized study is then to demonstrate how the asphericity behaves as a function of $N_\mathrm{fil}$, $j_\mathrm{fil}$ and $f_\mathrm{outflow}$. To this end we kept the total square integral in the filamentary component equal between different geometrical arrangements (i.e. we always assigned a total power $\int j_\mathrm{fil}^2$ to distribute over the $N_\mathrm{fil}$ filaments). The resulting maps can be found in \cref{varygrid}.
\begin{figure}
\centering
\includegraphics[width=\columnwidth]{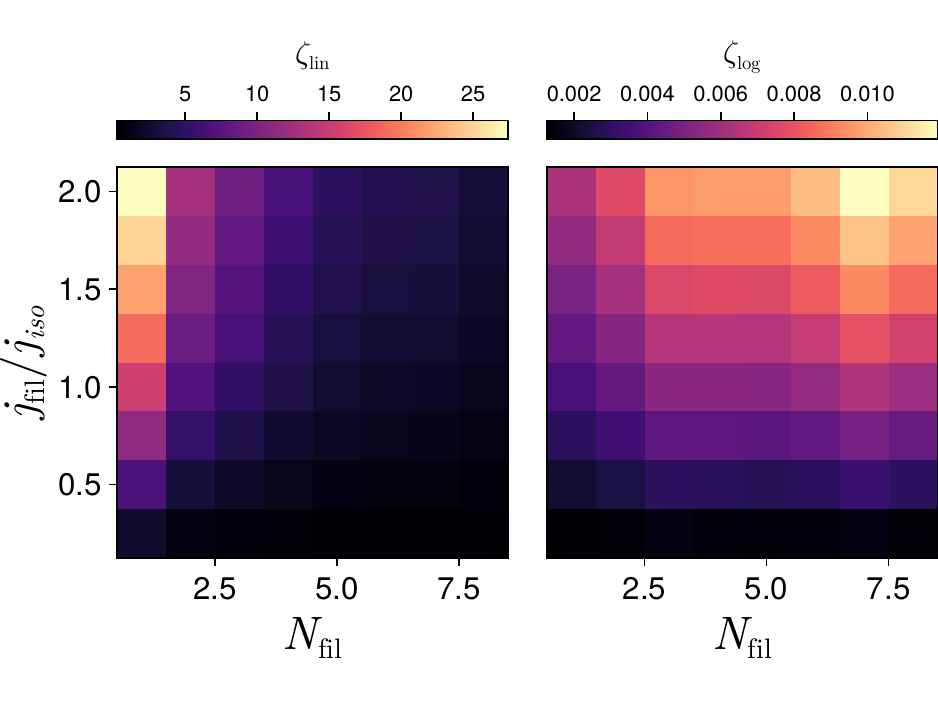}
\caption{Linear (left panel) and logarithmic (right panel) asphericities of the toy model. $f_\mathrm{outflow}$ is kept constant at 0.0. We vary the total number of filaments $N_\mathrm{fil}$ and the relative strength of filamentary inflow $j_\mathrm{fil}$ along the x and y axis respectively.}
\label{gridstrength}
\end{figure}

\begin{figure}
\centering
\includegraphics[width=\columnwidth]{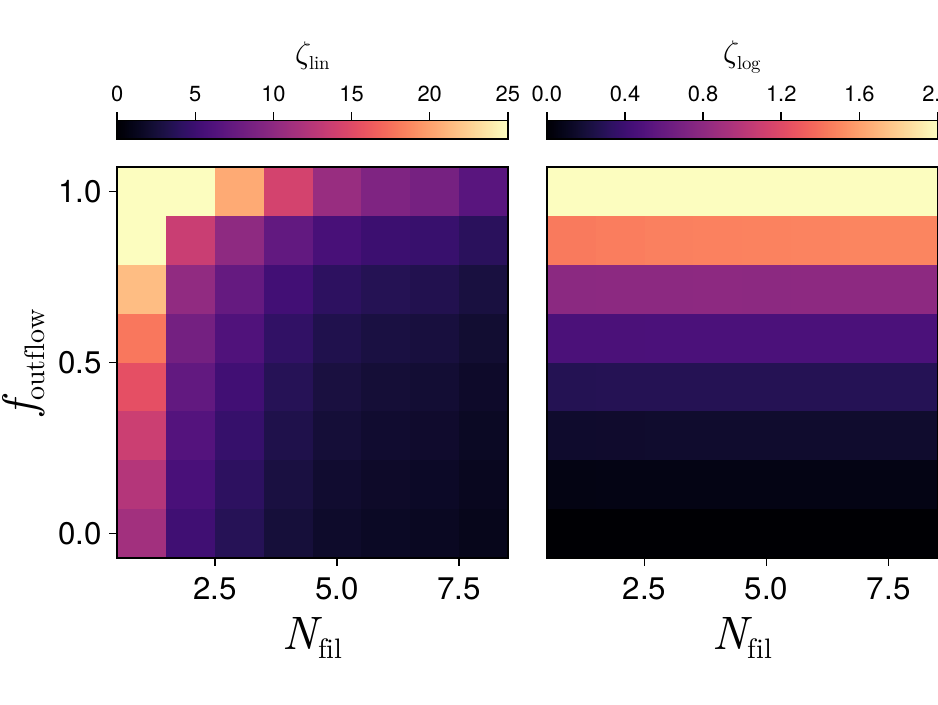}
\caption{Linear (left panel) and logarithmic (right panel) asphericities of the toy model. $j_\mathrm{fil}/j_\mathrm{iso}$ is kept constant at 3.0. We vary the total number of filaments and the outflow covering fraction along the x and y axis respectively.}
\label{gridzero}
\end{figure}

\begin{figure}[!htbp]
\centering
\includegraphics[width=\columnwidth]{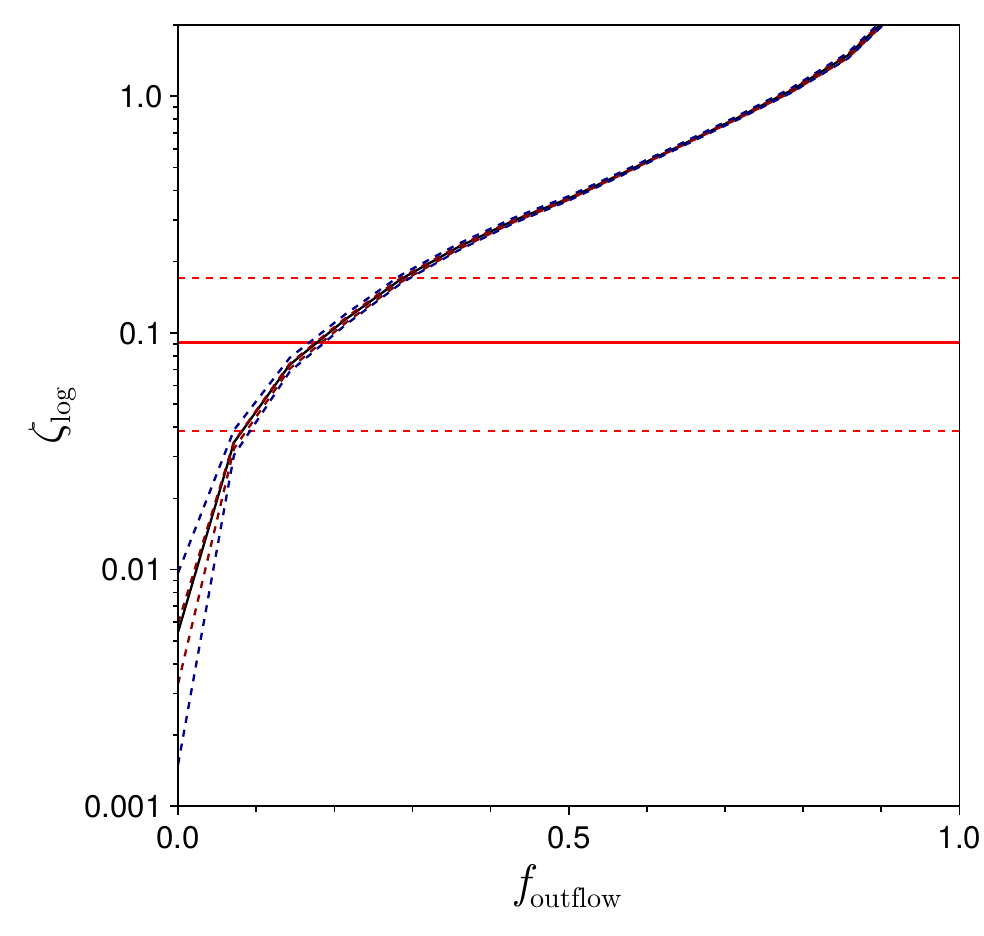}
\caption{Direct dependence of the logarithmic asphericity $\zeta_\mathrm{log}$ on the outflow covering fraction $f_\mathrm{outflow}$. The red lines mark the median (solid line) and 25th and 75th quantile (dashed lines) for galactic haloes in simulations at early times (see also \cref{aspherspace}). All other parameters of the model are kept constant ($j_\mathrm{fil}/j_\mathrm{tot}=1$, $N_\mathrm{fil}=4$). The dashed lines show the same functional dependence computed with the extremal model parameter values $N_\mathrm{fil}=1,8$ (dark red) and $j_\mathrm{fil}/j_\mathrm{iso}=0.25,2$ (dark blue).}
\label{omega}
\end{figure}
\cref{gridstrength} demonstrates the behaviour of the two flavours of asphericity when the filamentary strength $j_\mathrm{fil}$ and number of filaments $N_\mathrm{fil}$ are co-varied. Each grid point represents the asphericity of a single synthetic map. The left panel shows the behaviour of the linear asphericity $\zeta_\mathrm{lin}$, while the right panel shows the logarithmic asphericity, $\zeta_\mathrm{log}$. As can be seen, the two parameters behave orthogonally to each other: The asphericity of the linear map is very sensitive to the number of filaments $N_\mathrm{fil}$, with a strong peak at a dipole map with a single inflowing hotspot. Since the integral of the filamentary inflow is kept constant in each row in the figure, adding additional filaments lowers the overall asphericity. The logarithmic variation on the other hand is only weakly sensitive to the number of filaments and peaks at maps with a large number of inflow hotspots. This counter-intuitive positive scaling is a consequence of the aforementioned normalization, which is applied in linear space. Increasing the total inflow in the aspherical components expectedly increases both parameters, with the logarithmic variant exhibiting a stronger relative dependence.

\cref{gridzero} shows the impact of regions of zero inflow (i.e. outflowing regions) with the number of filaments $N_\mathrm{fil}$. As in the previous figure, the left panel shows the linear asphericity and the right panel shows the logarithmic asphericity. While increasing the opening angle of the outflow cone doesn't drastically change the linear asphericity, the logarithmic asphericity is very sensitive to this due to the reduced dynamical range as can be seen by values being more than an order of magnitude larger than for the case without holes. This strong scaling is present for all filament configurations. 

To study this behaviour of the logarithmic asphericity in greater detail, we plot it against the covering fraction of the outflow in \Cref{omega} with the other model parameters kept constant. There is a very strong power-law scaling at low covering fraction, then a transition to a shallower slope at around $10\%$ covering fraction. In the intermediate range the slope is almost constant with the asphericity reaching values typically found for galactic haloes in Magneticum (see companion paper Seidel+2026b in prep.) at around $15-20\%$. In the limit of a covering fraction of 1 the logarithmic asphericity converges into the regime of the linear values $\simeq 1$, as the isotropic component is completely removed, leaving only filaments and outflow regions. As can be seen from the enveloping dashed lines, the shape of this dependency is largely independent of $N_\mathrm{fil}$ and $j_\mathrm{fil}/j_\mathrm{iso}$, indicating that indeed the logarithmic asphericity is largely driven by the covering fraction of the outflows.
\begin{figure*}
 \centering
\includegraphics[width=\textwidth]{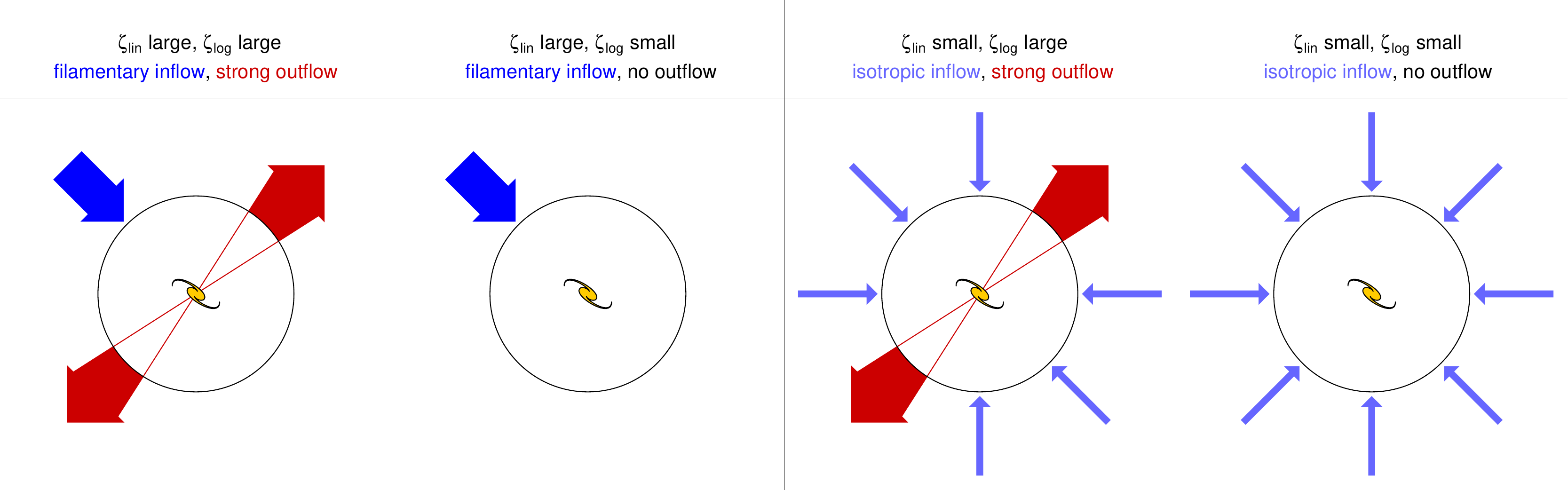}
\caption{Schematic presentation of four limiting cases with high and low $\zeta_\mathrm{lin}$ and $\zeta_\mathrm{log}$ respectively.}
\label{examp}
\end{figure*}
In summary, \cref{gridstrength} and \cref{gridzero} demonstrate that in combination $\zeta_\mathrm{lin}$ and $\zeta_{\log}$ trace to what degree the inflow is dominated by concentrated high-inflow regions such as filaments, as well as the fraction of the halo boundary that is actively inflowing and conversely the coverage of the outflow region. The findings of this section are summarized in sketch form in \cref{examp}: A high $\zeta_\mathrm{log}$ and $\zeta_\mathrm{lin}$ indicate a strongly anisotropic inflow focussed on few strongly inflowing filaments as well as an isotropic component that is strongly disrupted by outflows. A low $\zeta_\mathrm{log}$ in combination with a high $\zeta_\mathrm{lin}$ indicates a strongly focussed inflow with an unperturbed isotropic component. Conversely, if only $\zeta_\mathrm{log}$ is large, this is an indication of fairly isotropic inflow with multiple infall direction in combination with strong outflow components disrupting the weak isotropic component. The case with both parameters having low values represents the ideal case of a very isotropic and undisrupted inflow. 

These properties make the asphericity parameters ideal analysis tools to study how haloes interact with their environment both in terms of cosmological accretion and feedback from star formation and AGN activity.
\section{Tests on simulated flow fields}
\label{sec:Simtest}
\subsection{Simulations}
We used the Magneticum Pathfinder suite of simulations\footnote{www.magneticum.org} which features a large selection of simulation volumes and resolutions \citep{dolag2025}. All Magneticum Pathfinder simulations were performed with an updated version of the parallel cosmological Tree-PM code {\sc GADGET-2} \citep{springel2005}, namely {\sc P-GADGET3}. To simulate the hydrodynamics, a modified smoothed-particle hydrodynamics (SPH) scheme was employed, with modifications such as improved thermal conduction \citep{dolag2004} and a low-viscosity formulation of SPH \citep{dolag2005, beck2016a} specifically to improve the modelling of intra-cluster medium physics. Sub-grid physics include metal dependent cooling, star formation and detailed treatment of stellar evolution \citep{tornatore2003,tornatore2007} with supernova driven winds \citep{springel2003}, black hole growth and feedback from AGN \citep{springel2005a,fabjan2010,hirschmann2014}. 

Magneticum Pathfinder offers a number of different simulation runs, varying in size and resolution (and implemented physics). For the purpose of this work we used the largest high resolution box ($2\times2880^3$ particles), {\it Box2b/hr}. Past studies of galaxy cluster physics in this box include (but are not limited to): Studies of galaxy cluster substructures \citep{kimmig2023b}, studies of older, low-surface brightness galaxy clusters \citep{ragagnin2022}, galaxy populations and star formation quenching due to cluster environments \citep{lotz2019,lotz2021} and predictions for soft X-ray excess emissions from the outskirts of galaxy clusters and the filamentary warm-hot intergalactic medium (WHIM) \citep{churazov2023}. The evolution of proto-clusters from z=4 to the present day was investigated by \citet{remus2023}. 
We additionally used complementary data from {\it Box4/uhr} due to its higher resolution. The cosmological parameters of the runs used are: $h=0.704$, $\Omega_m=0.272$, $\Omega_{\Lambda}=0.728$, $\sigma_8=0.809$, $n=0.963$ consistent with the WMAP-7 cosmology \citep{komatsu2011}.  

\subsection{Reconstructing the peculiar mass flow from simulation data}
\label{sec:methods:pecflow}
The flow through a surface given a continuous velocity field $\vec{v(x)}$ can be generally computed by either integrating its divergence over the enclosed volume or via the Gaussian theorem by
\begin{equation}
\label{1}
\int_V \vec{\nabla} \cdot \vec{F} \,dv = \oint_{\partial V} \vec{F} \cdot \hat{n} \,da .
\end{equation} From a numerical perspective the latter method is the more sensible (integrating over a volume, scaling with $r^3$, vs. a surface, with $r^2$ scaling). The largest haloes in the sample reach several Mpc in virial radius or in other words scales where the expansion of the background cosmology becomes relevant, therefore we use \textit{peculiar} velocities throughout this work.
In SPH the gas discretisation is based on representative point particles that are smoothed over a volume of space with a density dependent kernel. To accurately model the gas flow given a set of gas \enquote{particles} it is therefore necessary to take into account the geometry and smoothing lengths of the kernels. This ensures that there are no holes in the density/velocity field due to the SPH particles' clustering leaving empty pixels, which in turn ensures that the holes in the inflow map actually correspond to real outflowing regions instead of holes in the density field at the surface. 
We use the program {\sc SMAC} \citep{dolag2005a} to generate flow maps from the simulations. The procedure is identical to the one described by \citet{seidel2025}. Particles are interpolated on a Healpix map using a kernel weighted interpolation scheme.
To demonstrate how the spherical harmonics decomposition works in practice, we applied the formalism to two example haloes from the {\it Box2b/hr} Magneticum simulations. 

The haloes are selected for their dynamical states, we deliberately chose one relaxed galaxy cluster and one dynamically active cluster following \citet{kimmig2023b} (ID 5 and ID 20 respectively) at z=0.25. We used the SMAC method to compute maps from the particle data of the simulation and measure the flow field at 1$r_\mathrm{vir}$ for each halo.
 \begin{figure}[!htbp]
\centering
\includegraphics[width=\linewidth]{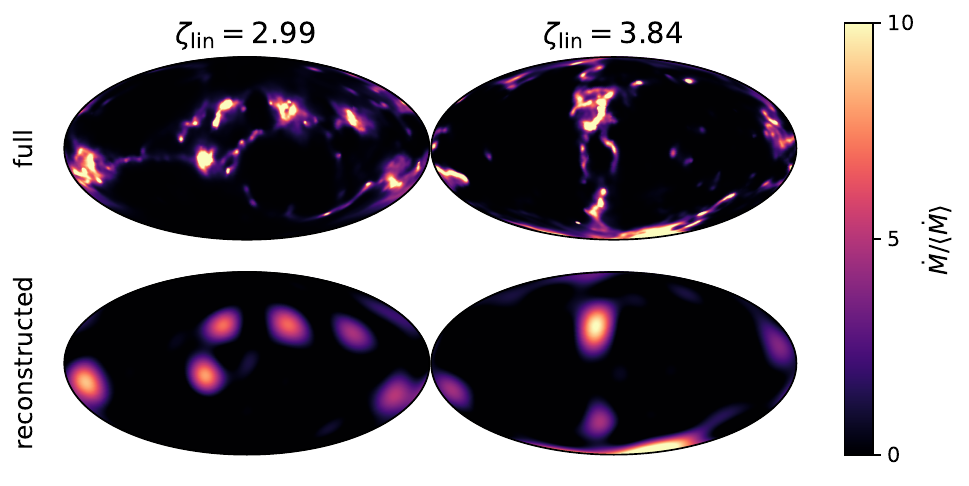}
\caption{Visualization of the effect of the spherical harmonics decomposition using the inflow field of two simulated haloes from Magneticum. The top row shows the full linear inflow field normalized by the mean inflow across the map. The dynamically active cluster ID 20 is shown in the left column, while the more relaxed ID 5 is shown in the right column. The bottom row shows the same map expressed only with spherical harmonics of order $l=1$ to $l=10$, filtering out the monopole and higher-order harmonics.}
\label{power}
\end{figure}
 \begin{figure}[!htbp]
\centering
\includegraphics[width=\linewidth]{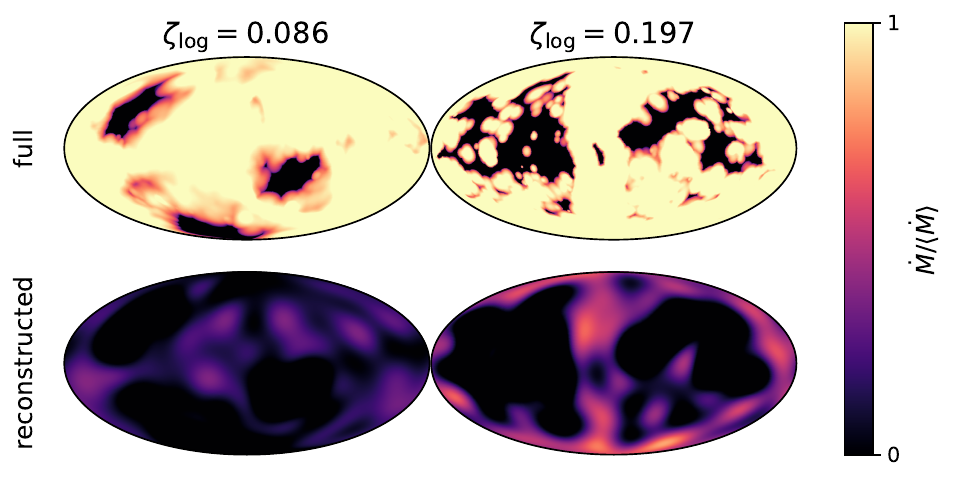}
\caption{Same as \cref{power} for the logarithmic flow field. The top row shows the full logarithmic inflow field normalized by the mean logarithmic inflow across the map. The bottom row shows the same map expressed only with spherical harmonics of order $l=1$ to $l=10$, filtering out the monopole and higher-order harmonics.}
\label{powerlog}
\end{figure}
\Cref{power} visualizes the obtained flow fields. It can be immediately seen from the linear inflow maps that while the toy model scenarios are simplifications, the decomposition into small circular high-flow field zones is a reasonable morphological approximation of the inflow these haloes receive, as the linear flow field is dominated by around 2-6 hotspots on the spherical surface. As for the test cases the halo with a greater number of distributed filamentary hotspots has a lower linear asphericity compared to the one with fewer filaments. 

In the bottom row we show the maps obtained by first computing the spherical harmonics decomposition and then setting the coefficients of all $a_{lm}$ with $l=0$ and $l>10$ to zero (essentially applying a bandpass filter). This demonstrates how the limits of the sum in \Cref{equation:asphericity} filter out the large-wavelength fluctuations in the flow field. The individual hotspots become even more apparent in the reconstructed maps with the smaller scale fluctuations filtered out. The dynamical state appears to not be directly connected to the instantaneous accretion field at z=0.25 since the values for both the relaxed and unrelaxed halo are similar. The fact that the dynamically more active cluster (ID 20) accretes more isotropically is in general consistent with the positive correlation between connectivity and dynamical state \citep{gouin2021}: As cluster 5 accretes through two major hotspots compared to the six inflow hotspots in cluster 20, its linear asphericity is higher because the flow varies on larger scales and is therefore less distributed over the virial surface (assuming a similar total inflow). For a complete picture of the connection between the dynamical state of galaxy clusters and their inflow asphericity, a temporal analysis is necessary, since the instantaneous flow state at the virial scale encodes the future of mass accretion onto the cluster itself rather than the past. We leave this thorough analysis for future work.

\Cref{powerlog} shows the same decomposition for the logarithmic maps. Computing the logarithm of the flow field significantly decreases the dynamic range of inflow values, emphasizing the regions with zero inflow (where the outflows dominate). This can be seen in the dark regions in the full maps. The spherical harmonics decomposition also reflects this morphology as the lower row shows. The second halo also has a higher logarithmic asphericity value, as a larger fraction of the spherical surface is covered by these zero-inflow regions. These two example haloes show that the behaviour of the two parameters in the test sample is indeed replicated in more complex flow fields. The spherical harmonics decomposition and the cut-off in $l-$degree effectively isolates the large scale fluctuations in the flow field corresponding to filamentary regions and infalling haloes. Additionally the decomposition of the logged flow field measures the large-scale fluctuation between inflow and outflow.
 \begin{figure}[!htbp]
\centering
\includegraphics[width=\linewidth]{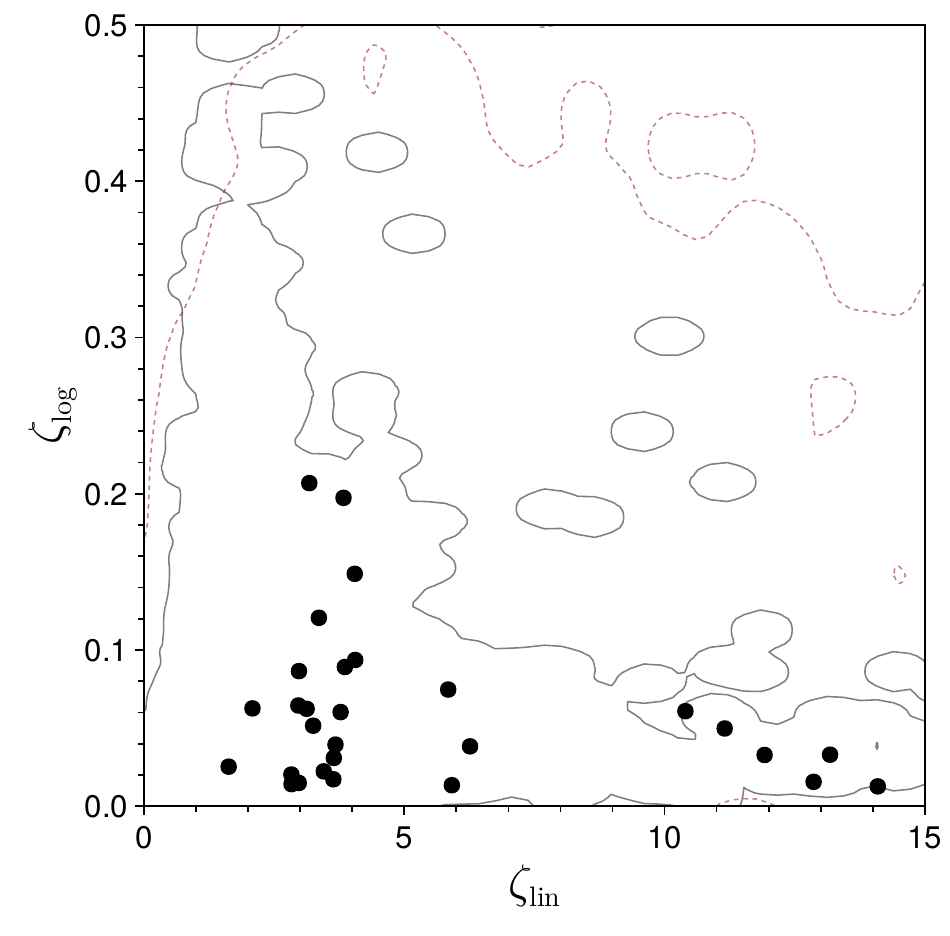}
\caption{Distribution in the asphericity-space of the 29 most massive galaxy clusters in Box2b (black circles) at z=0.25. The contours show a larger sample of galaxies with $M_\star>5\times 10^9M_\odot$ drawn from the higher resolution Box4/uhr simulation at z=0.25 (black) and z=1.3 (dark red dashed).}
\label{aspherspace}
\end{figure}

To put these two case studies into a wider context, we show the parameters $\zeta_\mathrm{lin}$ and $\zeta_\mathrm{log}$ for the 29 most massive galaxy clusters  plotted against each other in \cref{aspherspace}. As can be seen from the figure, at z=0 the clusters sit at relatively low logarithmic asphericities, while for the linear parameter two distinct groups can be identified. One set clusters at low to intermediate values while a second group can be found at rather high linear asphericities. Due to the small number of galaxy clusters and the limited resolution of Box2b, we also compare this to a more general halo sample from the smaller Box4uhr, where the higher resolution allows to compute the asphericities for a larger mass range. The black contours in \cref{aspherspace} show that the cluster sample is generally in good agreement with the larger sample, matching the striking L-shape of the contour. Overall this shows that at the present time, halo inflows are either dominated by a highly focused filamentary component (second column in \cref{examp}) or a strong outflowing component (third column). In contrast, the red contours show the parameter space to be much more evenly occupied at higher redshifts, where strong outflows and highly focussed inflows appear to coexist much more frequently. We study this temporal evolution of the morphology of inflow fields in greater detail in the companion paper to this work by Seidel et al. (2026b).


\section{Summary and Conclusions}
We introduced the inflow asphericity parameter, a new quantitative analysis tool for evaluating the inflow structure of haloes in numerical simulations. This tool is based on a spherical harmonics decomposition and measures how non-isotropic the inflow field around a given object of interest is. Using a toy model setup to simulate the inflow across a spherical boundary, we showed how the multipole decomposition works and which features this method extracts. For this we varied a number of components of the toy model, corresponding to filaments, isotropic inflow and outflowing regions. There are two complementary approaches to analysing the flow maps: Performing the spherical harmonics decomposition on the linear field extracts the high-flow regions where mass flows are orders of magnitude above the average mass inflow. On the other hand, using the logarithmic inflow field reduces the fluctuations in the inflow maps significantly, and the regions with zero inflow, where outflows dominate, become more important. 

In combination, these two parameters help distinguish galactic accretion in four distinct phases:
\begin{itemize}
\item Highly focussed, filamentary inflow in combination with a strong outflow. 
\item Filamentary inflow with an unperturbed isotropic component, no outflow.
\item Isotropic inflow, disrupted by outflow.
\item Unperturbed isotropic inflow.
\end{itemize}
These groups help identify how a given halo is coupled to the environment in numerical simulations. To demonstrate this, we applied the extraction method to two simulated haloes from the Magneticum simulations. We demonstrated how the features in these more complex flow fields were extracted by the formalism, and showed how it quantifies qualitatively visible differences in the flow fields of two distinct haloes from the simulations. By comparing the linear and logarithmic asphericities for an extended simulation sample, we furthermore demonstrated these two parameters to be uncorrelated, encoding different information about the flow field.

This makes the asphericity a powerful tool to analyse environmental coupling on the intermediate scale just outside of collapsed haloes. With its two variants, this parameter is able to reliably trace the inflow from the cosmic web as well as outflows from the halo interior. In the companion paper to this work, we further analyse how these two quantities are connected to galaxy properties like star formation rate, gas fractions and metallicities. We will further test how sub-grid modelling influences the flow field geometry around galactic haloes specifically.
\begin{acknowledgements}
BAS acknowledges support by the grant agreements ANR-21-CE31-0019 / 490702358 from the French Agence Nationale de la Recherche / \emph{Deut\-sche For\-schungs\-ge\-mein\-schaft, DFG\/} through the LOCALIZATION project. The Magneticum simulations were performed at the Leibniz-Rechenzentrum with CPU time assigned to the Project \textit{pr83li}. this work was supported by the DFG under Germany's Excellence Strategy EXC-2094\,--\,390783311. We are especially grateful for the support by M.~Petkova through the Computational Center for Particle and Astrophysics (C2PAP). LMV acknowledges support by the German Academic Scholarship Foundation (Studienstiftung des deutschen Volkes) and the Marianne-Plehn-Program of the Elite Network of Bavaria. LCK acknowledges support by the DFG project nr. 516355818. BAS and KD acknowledge support by the COMPLEX project from the European Research Council (ERC) under the European Union’s Horizon 2020 research and innovation program grant agreement ERC-2019-AdG 882679.
\end{acknowledgements}
\bibliographystyle{aa} 
\bibliography{bib.bib}
\begin{appendix}
\section{Special Cases}
\begin{figure}[!htbp]
\centering
\includegraphics[width=\linewidth]{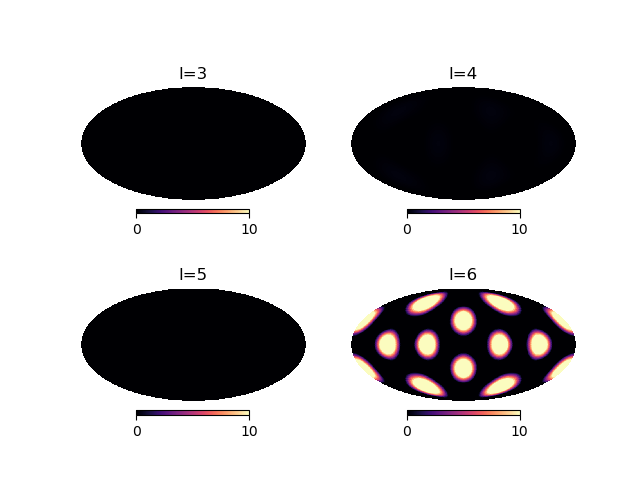}
\caption{Reconstruction of the Icosahedron setup with the spherical harmonics from $l=3$ to $l=6$.}
\label{Icodecomp}
\end{figure}
The tests in \cref{sec:testcases} are setup with the simplest geometry: A planar configuration with all filaments in one plane. In terms of the spherical harmonics, this corresponds to the terms with $|m|=l$, the so-called sectoral harmonics. We additionally tested the behaviour of the asphericity using more complex geometries, making use of the full 3D information encoded in the flow maps. 
\paragraph{Polyhedral configuration}
\label{platonics}
The simple planar setup corresponds to only a small subset of the spherical harmonics functions. In order to test more complex configurations we additionally constructed a set of symmetrical configurations derived from the five so-called platonic solids. These regular polyhedra are relatively simple to construct and provide a more general test case for infall geometry configurations. Following the procedure from the previous section we set up circular regions with a Gaussian profile at each vertex of each platonic solid. This provided five additional maps with 4,6,8,12 and 20 regions for the Tetrahedron, Octahedron, Cube, Icosahedron and Dodecahedron respectively. For the platonic solid setups, the simple correspondence between number of inflow regions and $l$ degree with maximum power breaks down because modes with $l\neq m$ play a role for these symmetries. As a specific example, the Icosahedron configuration with 12 vertices can be completely represented by a linear combination of the $m$ modes with degree $l=6$ (see \cref{Icodecomp}). This makes these setups an ideal test case for the behaviour of the asphericity parameters in more complex geometries. 
\subsection{Impact of the configuration on the asphericity parameter}
To ensure that the asphericity parameter is not affected by any specific geometry or choice of filament width, we test the simple relationship between filament number and asphericity for a number of different configurations. Additionally the possible impact of resolution is studied, since the spatial smoothing lengths can significantly impact the spherical harmonics decomposition.
\begin{figure}[!htbp]
\centering
\includegraphics[width=\linewidth]{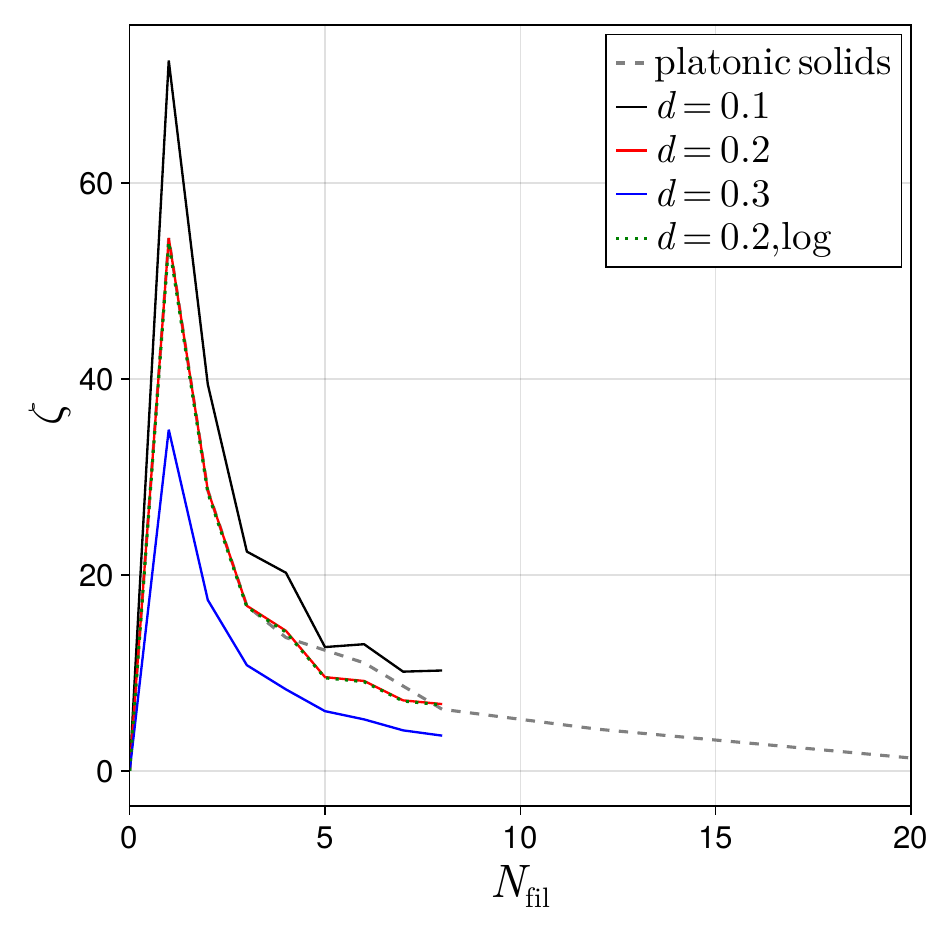}
\caption{Dependence of the linear asphericity on the Number and thickness of inflow hotspots in the test setup. The coloured curves from black to blue show the asphericity with increasing hotspot width. $j_\mathrm{fil}/j_{iso}$ is kept constant at 6/8 and $f_\mathrm{out}=1.0$. The grey curve shows the same asphericity continued by the platonic solid setups (see previous section). Additionally the green dotted line shows the logarithmic asphericity, which is identical to the linear asphericity at $f_\mathrm{out}=1.0$}
\label{Nvszeta}
\end{figure}
\section{Truncation of the sum}
\label{appendix:aggregate}
\begin{figure}[!htbp]
\centering
\includegraphics[width=\linewidth]{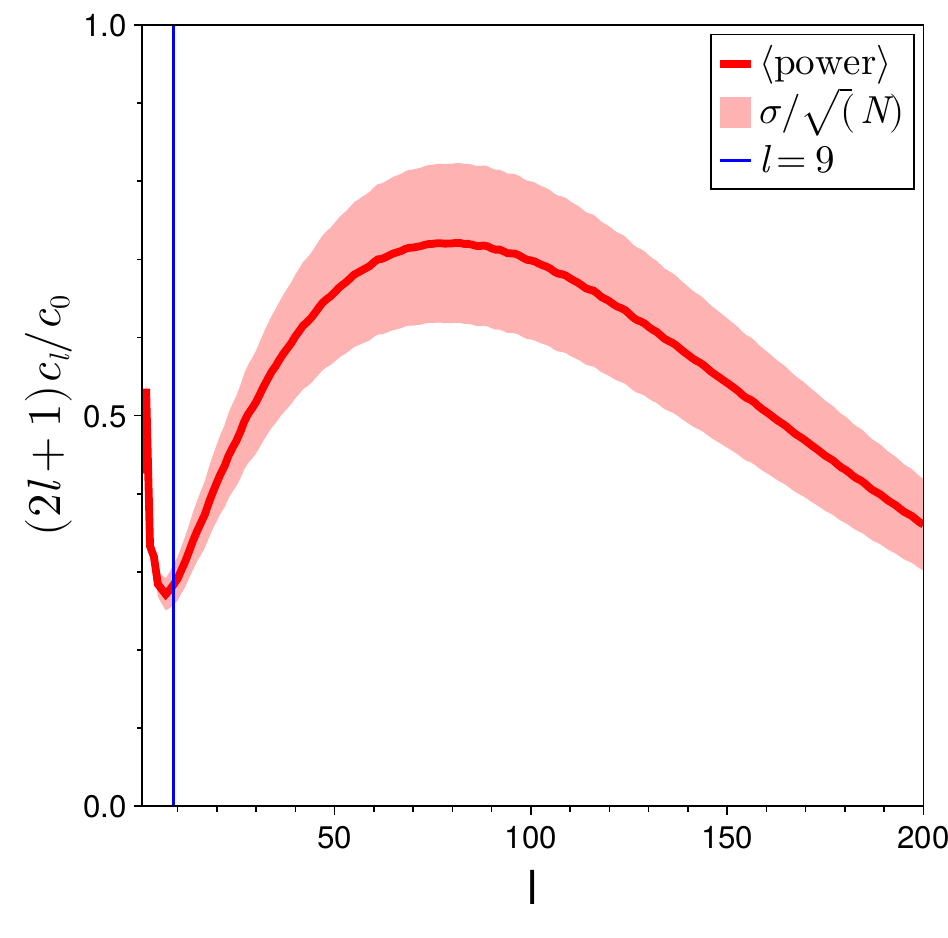}
\caption{Mean power spectrum of 2000 haloes from Box4. The shaded area shows the standard error of the curve, while the blue vertical line marks the low-pass cut-off point for the large-scale part of the spectrum chosen throughout this work.}
\label{aggspec}
\end{figure}

To motivate the choice of cut-off degree ($l=9$) for the computation of the asphericity parameter, \cref{aggspec} shows the combined spectrum for 2000 simulated haloes from Magneticum at z=0. The curve shows a characteristic dip at this specific wave number transitioning to a high-frequency rise at higher angular frequencies. 
\section{Spherical harmonics and filament connectivity}
\label{appendix:connectivity}
\begin{figure}[!htbp]
\centering
\includegraphics[width=\linewidth]{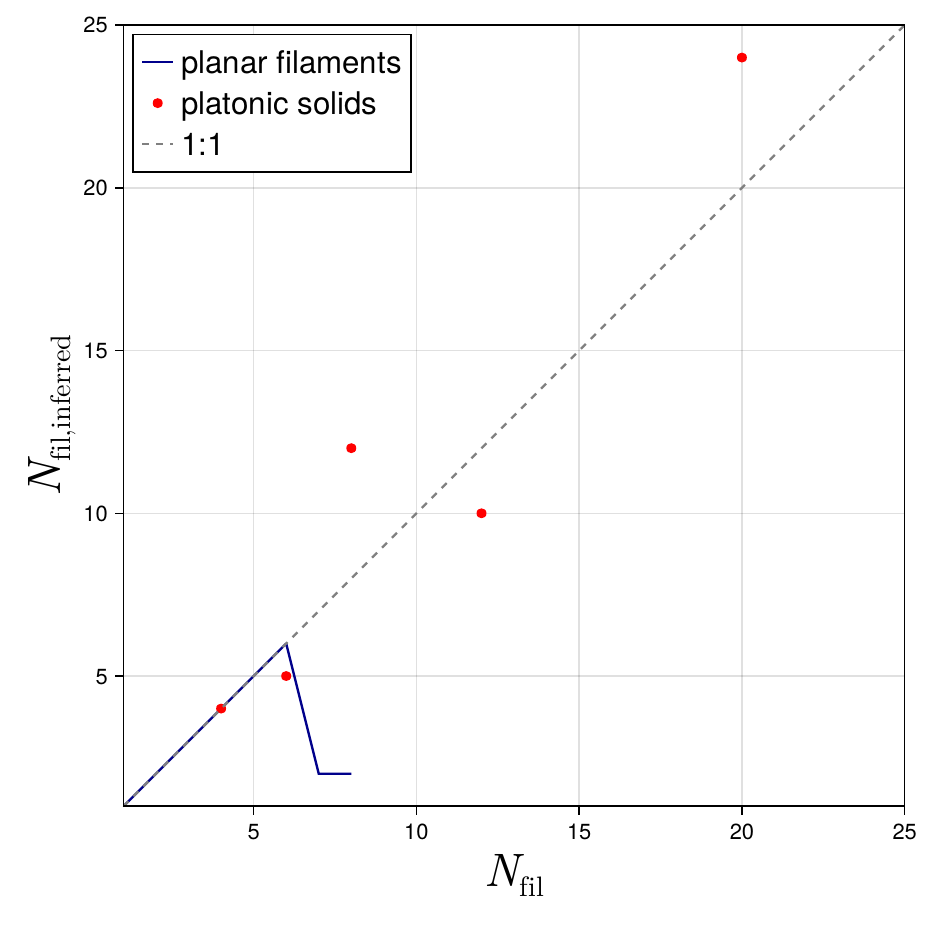}
\caption{Power spectra of the whole suite of idealized test cases with 1-8 hotspots.}
\label{powerspectest}
\end{figure}

From the value of the asphericity parameters itself, it cannot be conclusively derived how many filaments are feeding a specific system. This raises the question of whether this can be inferred from the full spectrum of $a_{lm}$. A first order approximation is to use the $l$ mode where the power spectrum of a given map is maximal. \cref{powerspectest} shows that (blue line) this works for the planar test maps (up to order 7, where the individual spots essentially form a single band, leading to a dipole -like power spectrum). For the platonic solid setups, this simple approach is no longer valid because morphologically these configurations are much more similar to tesseral harmonics with zones at multiple latitudes and longitudes. For tesseral spherical harmonics with $0<m<l$, the $2m$ meridian zero lines and the $l-m$ parallel isolatitude zero lines \citep[e.g.]{whittaker2021} separate the spherical surface into $2m(l-m+1)$ zones, half of which correspond to a peak in the flow field. Therefore an estimate of the number of peaks in a given map can be obtained from the $l,m$ of the maximal $|a_{lm}|$ of the decomposition.
\section{Calibrating the absolute values for normalization}
In order to configure the numeric values of the test maps in a realistic manner instead of setting arbitrary values, we used the simulations as calibration for these test maps. To that end, we computed a cumulative log histogram from the pixels of the inflow maps of 100 randomly chosen haloes. We then normalized the inflow values with the halo masses to account for the mass-inflow relationship \citep[see][]{seidel2025}. 
\begin{figure}[!htbp]
\centering
\includegraphics[width=\linewidth]{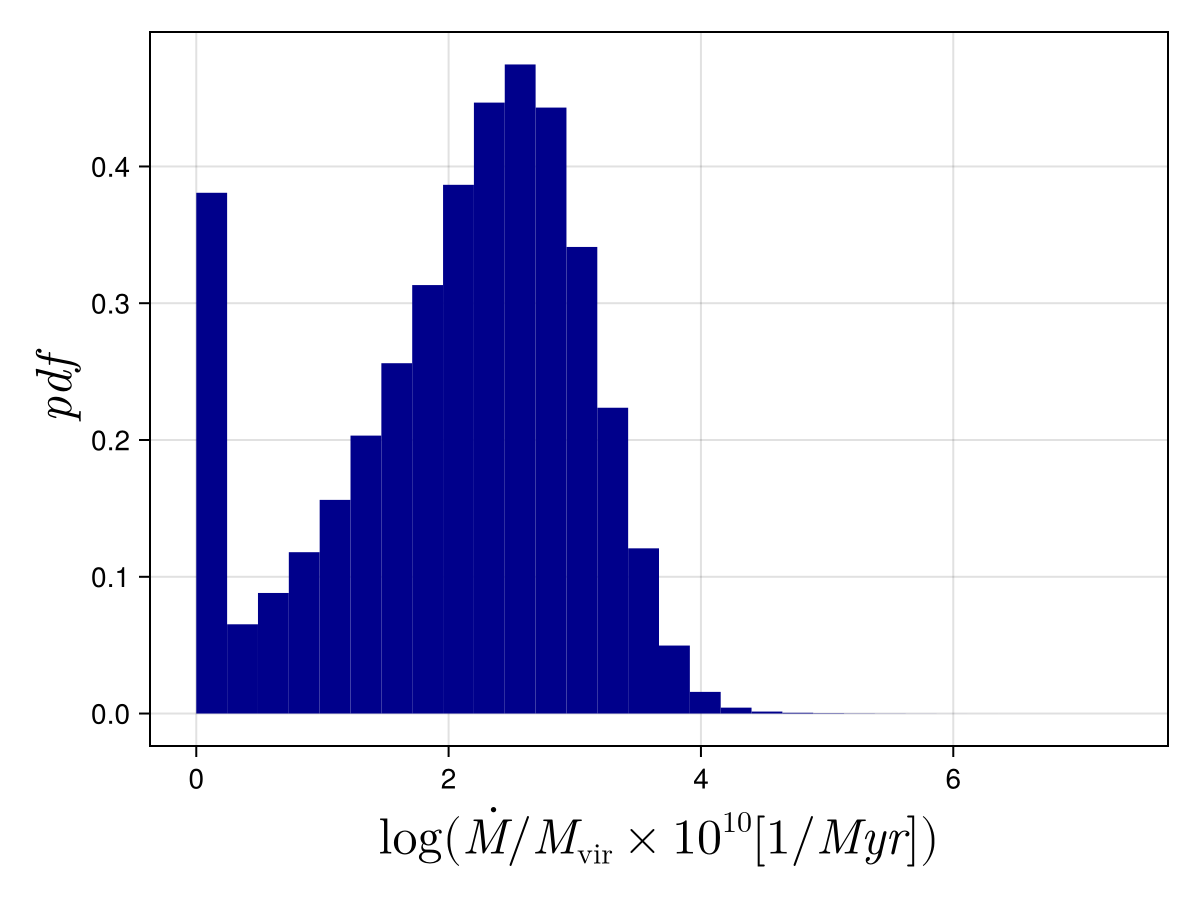}
\caption{Stacked histogram of inflow pixels for 100 haloes at z=0.}
\label{calib}
\end{figure}

\section{Variation of the parameters}
\begin{figure}[!htbp]
\centering
\includegraphics[width=\linewidth]{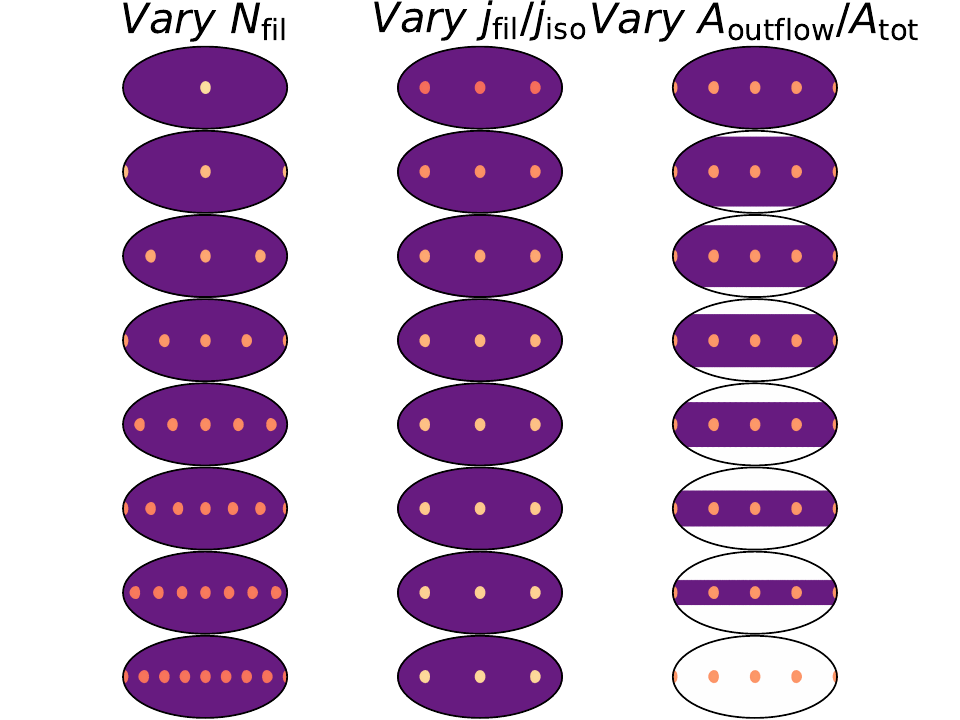}
\caption{Visualization of the three parameter variations for the toy model grids in \cref{gridstrength} and \cref{gridzero}.}
\label{varygrid}
\end{figure}
\cref{varygrid} qualitatively shows the effect of varying the different model parameters. The colour shows the synthetic flow rate with white indicating zero inflow in the last column.

\section{Orthogonality of the Spherical Harmonics}
Since the spherical harmonics decomposition we perform in this work is computed numerically, it is worthwhile to check how well this approximation holds for the inflow maps from the simulations but even more so the synthetic test map set. A good measure for the convergence of the numerical decomposition is the total power conservation: The spherical harmonics decomposition, similarly to i.e. a Fourier decomposition should obey a Parseval identity:
\begin{equation}
\int f^2dA\stackrel{!}{=}\sum (2l+1)c_l,
\end{equation}
namely the total power of the function is expected to be conserved in the power spectrum of the decomposition. Effectively this is a measure for the orthonormality of the numerically computed harmonics. Additionally, since the numerical decomposition is computed up to a truncation order $l_\mathrm{max}$, which is usually set to 2-3 NSIDE, the power conservation also depends on the completeness of the orthonormal basis.  
\begin{figure}[!htbp]
\centering
\includegraphics[width=\linewidth]{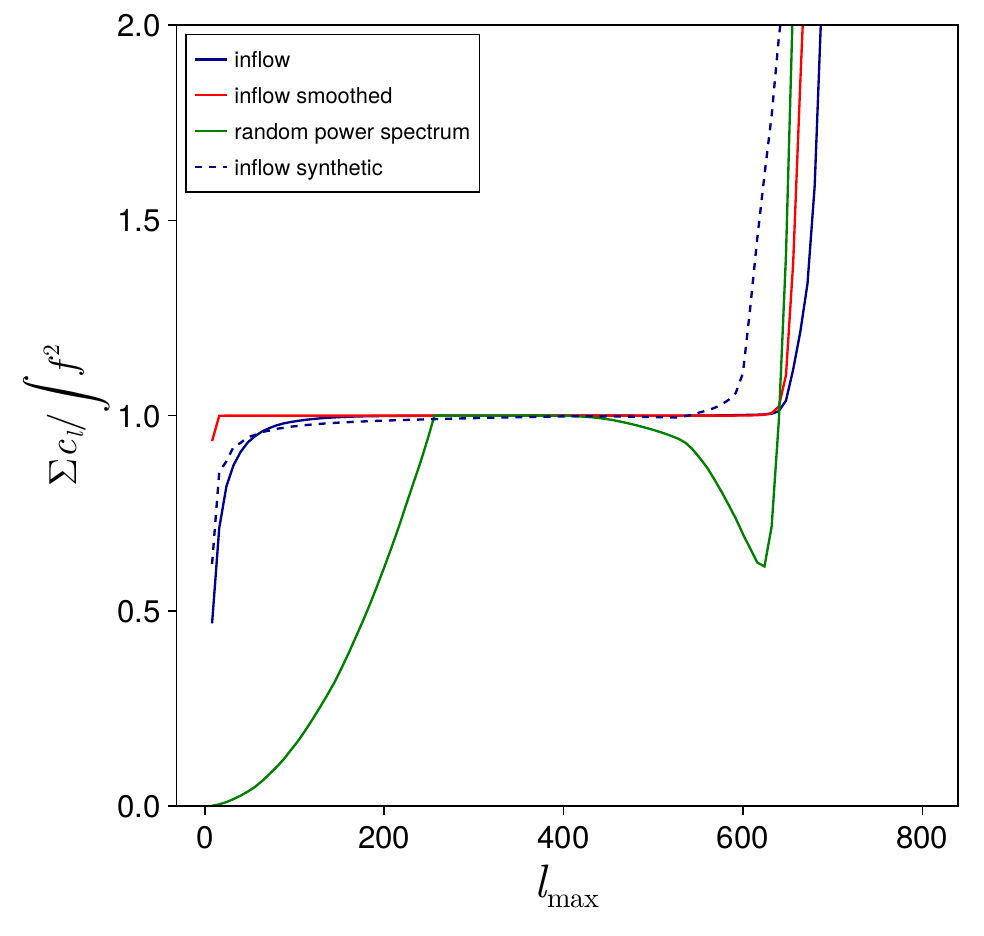}
\caption{Scaling of the Parseval Identity for spherical harmonics with the truncation l-mode of the decomposition. Shown is the scaling behaviour for a simulated inflow field (blue), a smoothed simulated field (red), a test map generated from spherical harmonics with randomized coefficients (green) and an example synthetic inflow map from \cref{Sec:Methods}. The pixel resolution of all maps is kept at NSIDE=128.}
\label{parseval}
\end{figure}
\cref{parseval} shows this scaling for four different map types, two obtained from simulated maps, one synthetic map from \cref{sec:methods:pecflow} and a reference map created by realizing a randomized power spectrum of spherical harmonics. As expected, all maps exhibit a "safe" region, where the Parseval identity holds well, which contains the truncation order employed throughout this work. After exceeding around 5 NSIDE, the decomposition breaks down as the frequency of the spherical harmonics approaches the sampling frequency of the healpix maps. More interestingly, at the low frequency end, both the simulated inflow maps as well as the synthetic maps show a much faster completeness, reaching identity at much smaller $l_\mathrm{max}$ than a map with a randomized power spectrum would. Since the synthetic maps scale almost identically to the simulated maps, they can be expected to be a faithful representation of the flow fields, validating the tests performed throughout this work.
\end{appendix}
%
%

\end{document}